\documentclass[10pt,conference]{IEEEtran}
\IEEEoverridecommandlockouts
   
   \usepackage{dblfloatfix}  
\usepackage{amsmath,amssymb,amsfonts}
\usepackage{graphicx}
\usepackage{textcomp}
\usepackage{xcolor}
\usepackage{adjustbox}
\usepackage{colortbl}

\usepackage{pgfplots}

\pgfplotsset{compat=1.10}
\usepgfplotslibrary{fillbetween}
\def\BibTeX{{\rm B\kern-.05em{\sc i\kern-.025em b}\kern-.08emT\kern-.1667em\lower.7ex\hbox{E}\kern-.125emX}}

\usepackage{array}    
\usepackage{booktabs} 
\usepackage[skins]{tcolorbox}
\usepackage{tabularx}
\usepackage{multirow}
\usepackage[flushleft]{threeparttable}
\usepackage{amssymb}
\usepackage{enumitem}
\usepackage{rotate}
\usepackage{graphicx}
\usepackage{colortbl}
\usepackage{arydshln}
\usepackage{times}
\usepackage{rotating}
\usepackage{makecell} 
\usepackage{tabularx}
\usepackage{balance}
\usepackage{booktabs}
\usepackage{wrapfig}
\usepackage{tikz}
\usetikzlibrary{angles}
\usepackage{makecell}
\usepackage{tabu}
\newcommand{\bi}{\begin{itemize}}
\newcommand{\ei}{\end{itemize}}
\newcommand{\be}{\begin{enumerate}}
\newcommand{\ee}{\end{enumerate}}
\newcommand{\fig}[1]{Figure~\ref{fig:#1}}

\newcommand{\tion}[1]{\S\ref{sect:#1}}
\usepackage{amsmath}
\usepackage{subfigure}

\usepackage{url}

\usepackage{tikz}
\usetikzlibrary{arrows.meta}
\tikzset{%
  >={Latex[width=2mm,length=2mm]},
            base/.style = {rectangle, rounded corners, draw=black,
                           minimum width=2.5cm, minimum height=1cm,
                           text centered, font=\sffamily},
  activityStarts/.style = {base, fill=blue!30},
       startstop/.style = {base, fill=red!30},
    activityRuns/.style = {base, fill=green!30},
         process/.style = {base, minimum width=2.5cm, fill=orange!15,
                           font=\ttfamily},
}

\makeatletter
\def\BState{\State\hskip-\ALG@thistlm}
\makeatother

\newcommand\MyBox[2]{
  \fbox{\lower0.75cm
    \vbox to 1.7cm{\vfil
      \hbox to 1.7cm{\hfil\parbox{1.4cm}{#1\\#2}\hfil}
      \vfil}%
  }%
}

\usepackage[final]{listings}
\usepackage{amsmath}
\usepackage{algorithm}
\usepackage[noend]{algpseudocode}

\usepackage[framemethod=tikz]{mdframed}
\usetikzlibrary{shadows}
\usepackage{graphics}
\newmdenv[
tikzsetting= {fill=gray!10},
linewidth=1pt,
roundcorner=2pt, 
shadow=false
]{myshadowbox}

\newenvironment{result}[2]
{\begin{myshadowbox}\textbf{\textit{\underline{Conclusion \##1:}}} #2}{ 
\end{myshadowbox}}

\definecolor{light-gray}{gray}{0.90}
    
\begin{document}
\newcommand{\IT}{SURVEY}

\title{  Better   Technical Debt Detection  via ``SURVEY''ing}

\author{\IEEEauthorblockN{Blinded for review}}

\author{\IEEEauthorblockN{Fahmid M. Fahid}
\IEEEauthorblockA{Computer Science \\
North Carolina State University\\
Raleigh, USA \\
ffahid@ncsu.edu}
\and
\IEEEauthorblockN{Zhe Yu}
\IEEEauthorblockA{Computer Science \\
North Carolina State University\\
Raleigh, USA \\
zyu9@ncsu.edu}
\and
\IEEEauthorblockN{Tim Menzies}
\IEEEauthorblockA{Computer Science \\
North Carolina State University\\
Raleigh, USA \\
timm@ieee.org}}


\maketitle
\thispagestyle{plain}
\pagestyle{plain}

\begin{abstract}
Software analytics can be improved by surveying; i.e.
 rechecking  and (possibly) revising
the labels offered by prior analysis.
Surveying is a time-consuming task and effective surveyors must carefully
manage their time. Specifically, they must balance the {\em cost} of further surveying
against the additional {\em benefits} of that extra effort. 

This paper proposes {\IT}0,
an incremental Logistic Regression estimation method that implements cost/benefit analysis.
Some classifier is used to rank the as-yet-unvisited examples according to how interesting
they might be. Humans then review the most interesting examples, after which their
feedback is
used to update an estimator for estimating how many examples are remaining. 

This paper evaluates {\IT}0  in the context of self-admitted technical debt.
As software project mature, they can accumulate ``technical debt'' i.e. developer decisions which are sub-optimal and decrease the overall quality of the code. 
Such decisions are often commented on by programmers in the code;
i.e. it is   self-admitted technical debt (SATD).  
Recent results show that text classifiers can automatically detect such debt. 
We find that we can significantly outperform prior results by {\IT}ing the
data. 
Specifically, for ten open-source JAVA projects, we can find 83\% of the technical debt via
{\IT}0 using
just 16\% of the comments (and if higher levels of recall are required, {\IT}0 can adjust towards that with some additional effort).
\end{abstract}

\begin{IEEEkeywords}
Technical debt,   software analytics
\end{IEEEkeywords}

\section{Introduction}
\label{section intro}
This paper is about cost-effective analytics using {\em surveying};
i.e. 
 rechecking  and (possibly) revising
the labels found by prior analysis.
We demonstrate the value of surveying by showing that, it can lead to   better
predictors for {\em technical debt}
than existing state-of-the-art methods~\cite{huang2018identifying}. 

Studying technical debt is important since it can significantly damage
project maintainability~\cite{cunningham1993wycash,guo2011tracking,nugroho2011empirical}.
When developers cut corners and make haste to rush out code, then that code
often contains {\em technical debt}; i.e. decisions that must be repaid, later on, with further work.
Technical debt is like dirt in the gears of software production.
As technical debt accumulates,   development becomes harder and slower. Technical debt can affect many aspects of a system including {\em evolvability} (how fast we can add new functionality) and {\em maintainability} (how easily developers can handle new or unseen bugs in code).


Surveying is important for automated software engineering
since many automated software analytics methods
assume that they are learning from correctly labelled examples.
However, before
an automated method uses labels from old data,   it is prudent to
revisit and recheck
the labels generated  by  prior analysis.
This is needed since humans often make mistakes in the labelling~\cite{hatton2008testing}. 
  But surveying can be  (very)
  time-consuming process. For example,
  later we show that surveying
  all the data used in this study would
  require more than 350 hours.
Clearly, surveying will not be apoted as standard
practice unless   we  can reduce its associated effort. 
  
  Algorithm~\ref{algo:zero} describes {\IT}0, a human-in-the-loop algorithm   for
  reducing the cost of surveying.
  The details of {\IT}0 are offered later in this paper.
For now, it is suffice to say,
 {\IT}0 includes an early exit strategy  (in Step5)
that is triggered
 if  ``enough'' examples have been found.
  
To assess the significance of {\IT}0, this paper builds predictors
for technical debt, with and without surveying. That experience let us answer the following research questions.

\begin{algorithm}[!b]
 \small
  \be   
  \item Randomly sort   $n$ software artifacts (e.g. code comments);
  \item Use prior data to build a classifier $C$ and a sorter $S$;
  \item Using $S$, ask reader $R$ to review  and label the first $m \ll n$ artifacts as ``good, bad'' (in our case, ``bad'' means ``has TD'');
  \item Using the labelled examples, update  an estimator $E$
  for the number of ``bad''  remaining in the  $n-m$ examples;
  \item Exit if $E$ says we found enough
  ``bad'' examples;
  \item Else:
   \bi
  \item Skip over the  first $m$ artifacts. Set $n=n-m$.
  \item Apply the sorter $S$ to arrange the 
  remaining artifacts,  in order
 of descending  ``bad''-ness; 
  \item
 Loop to Step3.
  \ei
  \ee
  \caption{  {\IT}0 = $\{C,S,E,R,m\}$. Using a human reader $R$, 
  a sorter $S$, a classifier $C$ and an estimator $E$, {\IT}0 updates its knowledge  every $m$ examples.}\label{algo:zero}
  \end{algorithm}

{\bf RQ1: Is surveying necessary?} 
If there are no disputes about what labels to assign to (e.g.) code comments,  there is no need for  surveying the data. But this is not the case. 
We find that labels  about technical debt from different sources
have many disagreements (36\% to 79\%, median to max).
Hence: 


\begin{figure*}[!b]

\begin{center}
\includegraphics[width=6in]{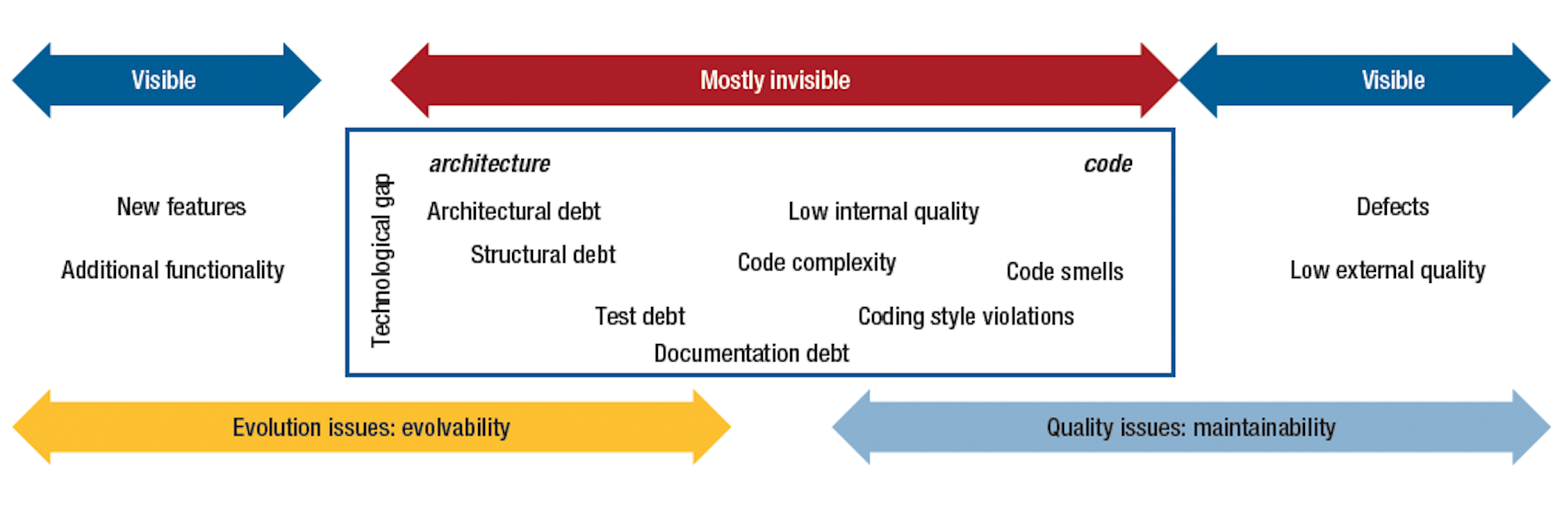}
\end{center}
\caption{Impact of technical debt on software. From~\cite{Ozkaya12}.}\label{fig:impact}
\end{figure*} 
\begin{result}{1}
Surveying is required to resolve
 disagreement about labels.
 \end{result}
 
{\bf RQ2: Is {\IT}0 useful?}  {\IT}0 cannot be recommended {\em unless} it
improves our ability to make quality predictions. 
Therefore we compared the predictive power of classifiers  that were
trained using {\IT}0's labels. We found that
{\IT}0's labels improved recall from 63\% to 83\% (median values across 10 data sets).
That is:
\begin{result}{2}
{\IT}0 improved quality predictions. 
\end{result}

{\bf RQ3: Is  {\IT}0 comparable to the state-of-the-art for human-in-the-loop AI?}
{\IT}0 is not  a research contribution {\em unless} it out-performs
other human-in-the-loop AI tools.
Therefore we compared our results to those
from an ``optimum'' tool.
For this study,  ``optimum'' was computed
by giving prior state-of-the-art   methods   an undue advantage
(allowing those methods to  
tune hyperparameter for better test results).
We found that:
\begin{result}{3}
{\IT}0 made its predictions
at a near optimum rate.
\end{result}

{\bf RQ4: How soon can   {\IT}0 learn quality predictors?}
{\IT}0 cannot be said to mitigate the relabelling problem
{\em unless} it finds most quality issues
using very few artifacts. Therefore we tracked how many artifacts {\IT}0 had to
show the humans before it finds most of the technical debts.
We found that {\IT}0 asked
   humans to read around  16\% of the comments while finding 83\% of the technical debts (median values across ten data sets).
\begin{result}{4}
   {\IT}0   can be recommended as a way to reduce the labelling effort.
\end{result}
{\bf RQ5: Can {\IT}0 find more    issues?}
The previous research question showed that {\IT}0 can find most issues after minimal human
effort. But  if finding (e.g.) 83\% of the quality issues is not enough,
can {\IT}0 be used to find  even more technical debt issues?
We find that {\IT}0's stopping rule can be modified to find more issues with additional cost of reading:
\begin{result}{5}
 {\IT}0 can be used to get additional desired level of quality assurance.
\end{result}

{\bf RQ6: How much does {\IT}0 delay human readers?}
Humans grow frustrated and unproductive when they wait for a system response for a long time.
Therefore we recorded how long humans had to wait
for {\IT}0's conclusions.
We found that {\IT}0 needs   half a minute  to find the   next   
$m=100$ most interesting 
programmer comments.
Humans, on the other hand,
need
 twenty minutes~\cite{maldonado2015detecting} to assess if those 100  comments  a for ``self-admitted technical debt'' (defined later in this paper).
That is, {\IT}0 delays humans by $0.5/(20+0.5) \approx 2\%$. To put that another way:
\begin{result}{6}
{\IT}0 imposed
negligible overhead  (i.e. less than 5\%) on the activity of human experts.
\end{result}
 The rest of this paper is structured as follows. In section~\ref{section: background}, we first discuss the background work and other related concepts that is needed for our study. In section~\ref{section methodology}, we give a brief description of our dataset, a detailed description of {\IT}0 and our experiment and evaluation methods. section~\ref{section results} discuss the results of our study. We discuss the threats to validity in Section~\ref{section threats to validity}. We will close our work    by discussing its implication and possible future directions.
 
 Note that, for reproduction purpose, all our data and scripts is publicly available (see  github.com/blinded4Review).

\section{Background}
\label{section: background}
\subsection{About Technical Debt}\label{sect:td101}
Technical debt (TD) effects
multiple aspects of the software development process (see \fig{impact}).
The term  was first introduced by Cunningham in 1993~\cite{cunningham1993wycash}. 
It is a widespread problem:
\bi
\item
In 2012, after interviewing 35 software developers from different projects in different companies, both varying in size and type, Lim et al.~\cite{lim2012balancing}
found  developers generate TD  due to factors like  increased workload,  unrealistic deadline in projects, lack of knowledge, boredom,
 peer-pressure among developers, unawareness or short-term business goals of stakeholders, and reuse of legacy or third party or open source code. 
\item
After observing five large scale projects, Wehaibi et al. found that, the amount of technical debts in a project may be very low (only 3\% on average), yet 
they create significant amount of defects in the future (and fixing such technical debts are more difficult than regular defects)~\cite{wehaibi2016examining}. 
\item
Another study on five large scale software companies revealed that, TDs contaminate other parts of a software system and most of the future interests are non-linear in nature with respect to time~\cite{martini2015danger}. 
\item
According to the SIG (Software Improvement Group) study of Nugroho et al., a regular mid-level project owes $\$8,57,500$ in TD and resolving TD has a Return On Investment (ROI) of 15\% in seven years~\cite{nugroho2011empirical}. 
\item
Guo et al. also found similar results and concluded that, the cost of resolving TD in future is twice as much as resolving immediately~\cite{guo2011tracking}. 
\ei
Much research tried to identify TD as part of code smells using static code analysis, with limited success~\cite{marinescu2010incode,marinescu2004detection,marinescu2012assessing,zazworka2013case,fontana2012investigating}. 
Static code analysis has a high rate of false alarms while imposing complex and heavy structures for identifying TD~\cite{tsantalis2011identification,tsantalis2015assessing,graf2010speeding,ali2012application}. 

Recently, much more success has been seen in the work on so-called ``self-admitted technical debt'' (SATD).
A significant part of technical debt is often ``self-admitted'' by the developer in code comments\cite{potdar2014exploratory}. 
In 2014, after studying four large scale open source software projects, Potdar and Shihab~\cite{potdar2014exploratory} concluded that developers intentionally leave traces of TD in their comments
with remarks like  like ``{\em hack, fixme, is problematic, this isn't very solid, probably a bug, hope everything will work, fix this crap}'' 
Potdar and Shihar et al. found 62 distinct keywords for identifying such TD~\cite{potdar2014exploratory}
(similar conclusions were made   by Faris et al.~\cite{de2015contextualized}).
In 2015, Maldonado et al. used five open source projects to manually classify different types of SATD~\cite{maldonado2015detecting} and found:
\bi
\item
SATD mostly contains requirement debt and design debt in source code comments; 
\item 
75\% of the SATD gets removed, but the median lifetime of SATD ranges between 18-173 days~\cite{maldonado2017empirical}. 
\ei

Another study tried to find the SATD introducing commits in Github using different features on change level~\cite{yan2018automating}. 
Instead of using a bag of word approach, a recent study also proposed word embedding as vectorization technique for identifying SATD~\cite{flisar2018enhanced}. 
Other studies investigated source code comments using different text processing techniques. 
For example, Tan et al. analyzed source code comments using natural language processing to understand programming rules and documentations and indicate comment quality and inconsistency~\cite{tan2007icomment,tan2012tcomment}. 
A similar study was done by Khamis et al~\cite{khamis2010automatic}. 
After analyzing and categorizing comments in source code, Steidl et al. proposed a machine learning technique that can measure the comment quality according to category~\cite{steidl2013quality}. 
Malik et al. used random forest to understand the lifetime of code comments~\cite{malik2008understanding}. 
Similar study over three open source projects was also done by Fluri et al.~\cite{fluri2007code}.

In 2017, Maldonado et al.   identified two types of SATD in 10 open source projects (average 63\% F1 Score) using Natural Language Processing (a Max Entropy Stanford Classifier) using only 23\% training examples~\cite{maldonado2017using}. 
A different approach was introduced by Huang et al. in 2018~\cite{huang2018identifying}. 
Using eight datasets, Huang et al. 
build Naive Bayes Multinomial sub-classifier for each training dataset using information gain as feature selection.
By implementing an ensemble technique on sub-classifiers, they have found an average 73\% F1 scores for all datasets~\cite{huang2018identifying}. 
A recent IDE for Ecliplse was also released using this technique for identifying SATD in java projects~\cite{liu2018satd}.

  To the best of our knowledge,  Huang et al.'s EMSE'18 paper
is the current state-of-the-art approach for 
identifying SATD.  Hence, we base our work
on their methods.


\subsection{About Surveying}
This section describes surveying, why
it is needed, and why cost-effective methods for surveying are required.

Standard practice in software analytics is for different researchers to try their methods on shared data sets.
For example, in 2010, Jureckzo et al.~\cite{Jureczko:2010} offered tables  of data that summarized dozens of open source JAVA projects.
That data is widely used in the literature. A search at Google Scholar on ``xalan  synapse'' (two of the Jureckzo data sets)
shows that these data sets are used in 177 papers and eight textbooks, 126 of which are in the last five years.

Reusing data sets from other researchers  has its advantages and disadvantages.
 One advantage is {\em repeatability of research results}; i.e. using this shared data, it is now possible and practical
to  repeat/repute/prove prior results. For examples of such kind on analysis, see the proceedings of the PROMISE conference
or the ROSE festivals (recognizing and rewarding open science in SE) at FSE'18, FSE'19. ESEM'19 and ICSE'19.  
See also all the lists of 678 papers which reuse data from the Software-artifact Infrastructure Repository  at Nebraska University (sir.csc.ncsu.edu/portal/usage.php).

Another advantage is {\em faster research}.
Software analytics data sets contain independent and dependent variables. For example,
in the case of self-admitted technical debt, the independent variables are the programmer comments and the dependent variable
is the label ``SATD=yes'' or ``SATD=no''.
Independent variables can often be collected very quickly
(e.g.  Github's API permits  5000 queries per hour).
However,
assigning the dependent labels is comparatively a much  slower task.
According to Maldonado and Shihab et al.~\cite{maldonado2015detecting}, classifying 33,093 comments as ``SATD $\in$ \{yes,no\}''
from five open source projects  took approximately 95 hours by a single person; i.e.  10.3 seconds per comment.
Using that information, we calculated that, relabelling the data used in this paper would require
months of work (see Table~\ref{tbl:costs}).
When a task takes months to complete,
it is not surprising that research teams tend to reuse old labels rather
than make their own.

That said, the clear disadvantage of reusing old labels is {\em reusing old mistakes}.
Humans often make mistakes when  labelling~\cite{hatton2008testing}.  Hence,
it is prudent 
to  
  review   the labels found in a dataset. We used the term ``surveying' to refer to the process of 
revisiting,  rechecking, and possibly revising
the labels offered by prior analysis

 In our experience, surveying is  usually done on a somewhat informal basis.
 For example, researchers  would manually survey
a small number of randomly selected
 artifacts
  (e.g.  1\% of the corpus; or 100 artifacts). There are many problems with the informal
  approach to surveying:
  \bi
  \item How many
  random selections are enough?   That is, 
  on what basis should we select?
   \item
  And
   when to stop surveying?
  Should finding $N_1$ errors prompt $N_2$ more samples? Or is there some
  point after which further surveying is no longer cost-effective? 
  \ei
  In order to answer these questions,
the rest of this article discusses cost-effective methods for surveying. 
 

\begin{table}[!t]
 \caption{Cost of labelling, Three different scenarios.}\label{tbl:costs}
\small
\begin{mdframed}[backgroundcolor=blue!5]

{\bf Scenario \#1) Hackathons:}
Our dataset contains $62,275$ comments (from ten projects). At 10.3 seconds/comment~\cite{maldonado2015detecting},
this   takea 178 hours to label.
With two readers (one to read, one to verify)  this time becomes 356 hours.
Using the power of pizza, we can assemble a hackathon team of half a dozen graduate students willing to work on tasks
like this, six hours per day, two days per month; i.e. 6*6*2=72 hours per month. 

~\\

{\bf Scenario \#2) Teams of Two:}
Note that, if pushed, we could demand more
time from these students. For example, we could demand that two students work on this task, full time.
Given the tedium of that task, we imagine that they could work productively on this task for 20 hours per week per person.
Under these conditions, revisiting and relabelling our data would take nearly two months.

~\\

{\bf Scenario \#3) Crowdsourcing:}
 Given sufficient funds, such labelling could be done at a much faster rate.
 Crowdsourcing tools like Mechanical Turk could be used to assemble any number of readers
 to revisit and relabel all comments, in just a matter of hours~\cite{chen19icpc,wang19ist,wang19tse}. While this is certainly a useful heuristic method for scaling up labelling, ever so often there must a validation study
 where the results of crowdsourcing are checked against some ``ground truth''.
 This paper is concerned with cost-effective methods for generating that ground truth.
 \end{mdframed}

 \end{table}
 \section{Related Work}
 The process we call {\em surveying} uses some technology from
 research on  {\em  active learning}~\cite{cormack2014evaluation,Abe:1998}. 
 ``Active learners'' assume that some {\em oracle} offers labels
 to examples and that there is a cost incurred, each time we
 invoke the oracle
 (in the case of surveying, that might mean asking a human to check if a particular code comment is an example of SATD).  
 
 The research of active learning was certainly motivating for this work.
 However, standard active learning methods were not immediately applicable
 to the problem of technical debt. Accordingly, we  made
 numerous changes to standard methods.

Firstly, 
 {\em {\IT}0's workflow are different (more informed)}
 than those of an standard active learner.
 Such learners do not   know when to stop learning. 
 Since our goal is to understand how many more items we need to read,
 {\IT}0   adds an  incremental estimation method
that studies how fast humans are currently finding interesting examples and imposes a stopping criteria based on that estimation.   That estimation method is described later in this paper.

(Aside: outside the machine learning literature,
we did find two information retrieval methods for predicting
when to stop incremental learning 
from 
Ros et al.~\cite{ros2017machine} and
Cormack~\cite{cormack2016engineering}. When we experimented
with these methods, we found that our estimators out-performed
these methods. For more on this point, see the {\bf RQ3} results
discussed later in this paper.)

Secondly, {\em we needed different learning methods}.
Active learning in SE has been applied previously in (e.g.) the code search recommender tools of
 Gay et al.~\cite{gay09}  that seek 
   methods   implicated in   bug reports.
Our work  is very different to that:
\bi
\item
 Code search recommender tools
  {\em  input} bug reports and {\em output} code locations. In between, those tools search {\em static code descriptors}; i.e. theirs are a   {\em code analysis} tool.
  \item
  The tools of this paper
\textit{input} programmer comments and \textit{output} predictors of   technical debt.
In between, our methods search {\em text comments};
i.e. ours is a
{\em text mining} tool.
\ei
Thirdly,
{\em we had to make more use of prior knowledge}:
\bi
\item
Initially, we tried tools built to help researchers find (say) a few dozen relevant papers
within 1000 abstracts downloaded from Google Scholar~\cite{yu2018finding}. Those tools were not successful
(they resulted in single digit recall values). 
\item
On investigation, we realized
those tools started  learning afresh for each new problem. That is, those tools assumed that prior knowledge was not
relevant to new projects.
\item 
That assumption seemed inappropriate for ths paper
since, for  most commercial software developers, software is more {\em extended} and 
{\em refined} than build from scratch.
In such an environment, it is possible to discover important lessons from prior projects.
\item
Hence, as shown in Step2 of Algorithm~\ref{algo:zero}, {\IT}0 {\em starts} by learning models
from all prior projects. After that, the rest of {\IT}0 uses feedback from the current
project to refine the estimations from those models.
\ei

\section{Inside {\IT}0}\label{section methodology}
 
\noindent
Recall from the above that {\IT}O is characterized by:
\[\{C,S,E,R,m\}\]
 That is, {\IT}0 updates its knowledge  every $m$ example, using  a  reader $R$, 
   a sorter $S$, a classifier $C$ and an estimator $E$. In the experiments of this paper, $m$ defines
 how much data is passed to humans (each time, we pass $m=100$ examples). 
 
 The rest of this section describes $C,S,E,R$.  Just to say the obvious,
 this section includes many engineering choices which future research
 may want to revisit and revise. We make no claim that {\IT}0 is the best
 surveying tool. Rather, our goal is to popularize the surveying problem
 and produce a baseline result which can be used to guide
 the creation of better surveyors.


\subsection{About the Classifiers ``$C$''}

This paper compares two classifiers
\bi
\item $C_1$ = ensemble decision trees (EnsembleDT)
\item $C_2$ = linear SVM
\ei
These learners  were selected as follows.
Firstly,
as to our use of SVMs, these are commonly used method for text mining~\cite{hearst1998support}. 
A SVM outputs a map of binary sorted data with margins between the two as far apart as possible, known as the support vectors. 
SVM uses a {\em kernel} to transform   problem data into a higher dimensional space where it is easier
to find  decision boundary between examples.
SVM models this boundary as set of {\em support vectors}; i.e. examples of different classes closest to the boundary.   Depending on the kernel used with a SVM, their training times can be very fast or very slow.
For this work, we used linear SVM and SVM with radial bias functions. There was no
significant performance delta between them and the linear SVM was much faster. Hence,
for this work, we are reporting linear SVM only.

Secondly, as to our use of EnsembleDT,
our aim was to extend the results of   Huang et al.'s EMSE'18 paper. 
That work used an ensemble Naive Bayes classifiers:
In that approach
\bi
\item
The authors first trained one Naive Bayes Multinomial (NBM) sub-classifier for each training project.
\item
These solo classifiers were then consulted as an {\em ensemble},
where each solo classifier voted on whether or
not some test example was an example of SATD. 
\item
The output of such an ensemble classifier is the majority vote across all the ensemble members.
\ei
To build their system, Huang et al.
 used   Weka      library (written in Java)\cite{witten1999weka} and their  built-in ``StringToWordVector'' for vectorization and the ``NaiveBayesMultinomial'' for classification. 
We were unable to find an equivalent vectorizer in Python, so we  used the standard
TF-IDF vectorizer. 
We failed to reproduce their results using Scikit-learn's~\cite{scikit-learn} Naive Bayes Multinomial. 
But by retaining the ensemble method (as recommend by Huang et al.)
and switching the classifier to Decision Trees (DT) , we obtained   similar results. 
Thus, for our experiment, we used their framework and   data (with 2 additional projects) but with a modification to
the leaner (Decision Trees, not Naive Bayes Multinomial).

 Decision tree learners recursivily split data such that
 each split is more uniform than its parent. 
 The attribute used to split the data
   is selected in order to minimize the diversity of the data
after each split.
This is a very fast and efficient machine learning algorithm~\cite{safavian1991survey}.


 
 \subsection{About the Sorters ``$S$''}\label{sect:sorter}

{\IT}0 asks its learners to sort examples by how ``interesting'' they are.
Our two classifiers need  different sorters:
 \bi
  \item
 For EnsembleDT, the  $S$ function counts how many times ensemble members vote for SATD. 
 \item  For linear SVMs, a ``most interesting'' example would be an unlabelled artifact
 on the SATD side of the decision boundary, and furthest away from that boundary.
 Hence, the sorter $S$ for linear SVMs is ``distance from the bounary''
 (and for this measure, we take the SATD side of the boundary to be
   positive distance).

 \ei
Note that when our estimator needs the probability
that an example is  technical debt,
we reach into these sort orders and report the position of that example
w.r.t. the other examples. Formally, those
probabilities are generated by normalizing the sort scores over the
range between 0..1


\begin{figure}[!t]
\begin{center}
\includegraphics[width=\linewidth]{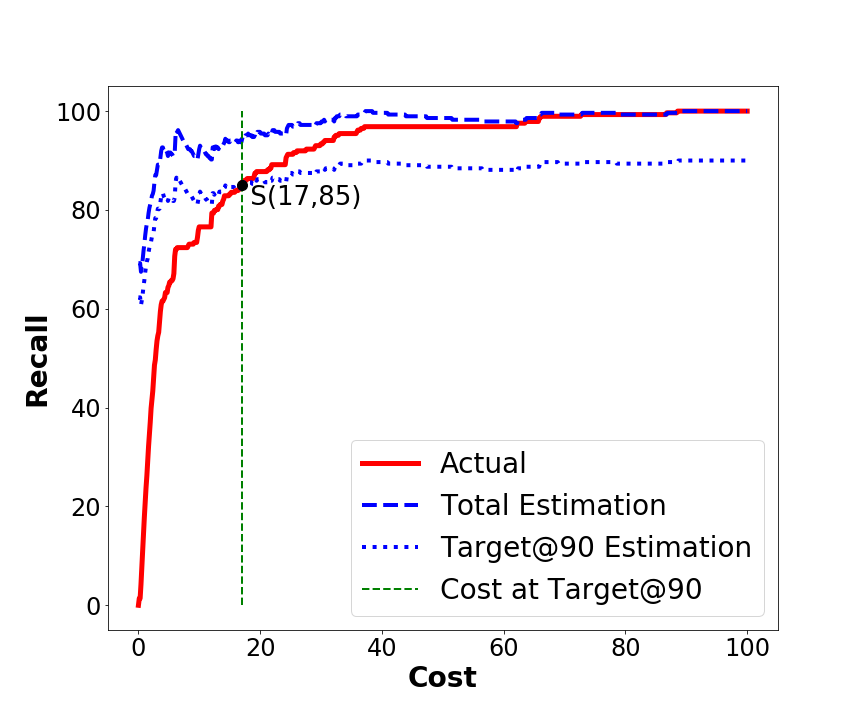}
\caption{Example retrieval curve (project SQL) using {\IT}0. ``Actual'' is the retrieval by the human according to the sorter $S$. ``Total Estimation'' is the output from the estimator. With Target@90, that becomes the ``Target@90 Estimation''. This intersects at point \textbf{S} where we stop with 85\% recall and 17\% cost.}
\label{fig framework example sql12}
\end{center}
\end{figure}

\subsection{About the Estimator ``$E$'' }

{\IT}0 uses an internal estimator, built using a Logistic Regression (LR) curve. 
Using this estimator, it is possible to   guess   how
many more interesting examples are left to find.

This estimator is used as follows. First, users specify a
{\em target} goal e.g. find 90\% of all the technical debt comments.
Next
{\IT}0  executes, asking the reader $R$ to examine $m=100$ comments at a time. As this process continues, more and more of the technical
debt comments are discovered. 

Figure~\ref{fig framework example sql12} shows a typical growth
curve. The dotted blue line shows the evolving estimator.
In practice, $E$ often over-estimates how much technical debt has been found. Hence, after reading $x=17\%$ of the code, the estimator reports that the target; i.e. that 90\% of the TD has been found (even though the exact
figure is 85\%, see Figure~\ref{fig framework example sql12}).

 Algorithm~\ref{algo:estimator} descibes our estimator.
 The estimator takes two inputs, some probability from
 the classifier (see \tion{sorter})
 and the labels. All unlabeled example are assumed to be ``not technical debt'' (because as shown in Table~\ref{table dataset details}, actual TD comments are quite rare). A logistic regression model is then  trained using the probabilities (from the learners) as
 the  independent variable; and labels as dependent examples. Using an iterative approach, the label for the unlabeled dataset is then predicted and the total number of remaining target class is calculated.

\begin{algorithm}[!t]
\small
\be
\item Count the total number of positives in $y_i$, say $C_i$.
\item Train a Logistic Regression curve $LR$ using $x$ and $y_i$;
\item Use $LR$ to predict the probabilities of $m_{u}$ (all the unlabeled datapoints), say $p_i$. 
\item Sort $p_i$ in decreasing order.
\item In each datapoint $m_{u_j} \in $ sorted $p_i$ not marked as ``seen'', calculate a cumulative sum of the
probabilities from the sorted list one by one and mark each datapoint as ``seen''. Whenever the sum $> 1$, 
reset the probability of the first one, $m_{u_j}=1$ and rest as 0. Go back to step 5 until all datapoint is marked as ``seen''. At the end of this step $p_i$ has new probabilities with $0$ and $1$s only.
\item Marge $p_i$ (new probabilities of unlabeled examples) with $m_l$ (labeled examples) and get $y_{i+1}$.
\item Count the total number of positives in $y_{i+1}$, say $C_{i+1}$.
\item If $C_i \neq C_{i+1}$, go back to step 1 with $y_{i+1}$ as new $y_i$
\ee
\caption{  {\IT}0 estimator with $m_{l}$ labeled examples by human (1 is SATD, 0 is Not-SATD) and $m_{u}$ unlabeled examples all marked with $0$ (because the dataset is very imbalanced). This is our $y_i$. 
The algorithm obtains its probabilities from the sorter $S$
described in \tion{sorter}. }\label{algo:estimator}
  \end{algorithm}

\subsection{About the Reader ``$R$'' }
{\IT}0 use a human expert to label examples in the test project. At each iteration, the sorter suggests $m$ most likely target class examples from the unlabeled data points and the human labels them one by one. 

In this experiment, we have implemented an automated human oracle to mimic the behaviour of a human reader. To do that, we kept the actual label of our test project (labeled by the authors of the data set~\cite{maldonado2015detecting}) as a separate reference set. At each iteration, the oracle looks into the reference set and label the comment (thus mimicking a human expert).

\section{Experimental Materials}
\subsection{Evaluation Metrics}

\textbf{Recall:} Our framework is concerned with how much target class (SATD) is found within the comments that has been checked. 
Formally, this is known as recall:

\begin{equation}
    \begin{aligned}
    Recall = \frac{True Positive}{True Positive + False Negative} \times 100\\
    = \frac{SATD Found}{SATD Found + Non SATD Found} \times 100 
    \end{aligned}
\end{equation}

The {\em larger} the recall,
the {\em better} retrieval process

\textbf{Cost:} As our framework has a human involved, we wanted to measure the cost of finding target class (SATD). 
For that, we only focused in the number of comments to read as a ratio of total number of comments. Thus, 

\begin{equation}
    \begin{aligned}
    Cost = \frac{Comments Read}{Total Comments} \times 100
    \end{aligned}
\end{equation}

Cost is a measurement of the overall effort needed for the human. 
The {\em smaller}  the cost the {\em better} the surveying.

\begin{table}[!t]
\small
\centering
\caption{Dataset Details. In ten projects,
the self-admitted technical debt comments are around 5\% of
all comments.}
\label{table dataset details}
\begin{tabular}{|p{0.4in}cp{0.3in}rrr|}\hline
   & 
  & 
  & 
\textbf{C = } &
\textbf{S = } & S/C\\
 & 
\textbf{Release} & 
\textbf{ ~} & 
\textbf{Comments} &
\textbf{SATD} &
\textbf{*100}\\
\hline
Apache Ant & 
1.7.0 & 
Automating Build&
4098 &
131 &
3.2\\
\hline
Apache JMeter & 
2.10 & 
Testing &
8057 &
374 &
4.64\\
\hline
ArgoUML & 
- & 
UML Diagram &
9452 &
1413 &
14.95 \\
\hline
Columba & 
1.4 & 
Email Client &
6468 &
204 &
3.15 \\
\hline
EMF & 
2.4.1 & 
Model Framework &
4390 &
104 &
2.37 \\
\hline
Hibernate Distribution & 
3.3.2 & 
Object Mapping Tool &
2968 &
472 &
15.90 \\
\hline
jEdit & 
4.2 & 
Java Text Editor &
10322 &
256 &
2.48 \\
\hline
jFree Chart & 
1.0.19 & 
Java Framework &
4408 &
209 &
4.74 \\
\hline
jRuby & 
1.4.0 & 
Ruby for Java &
4897 &
622 &
12.70 \\
\hline
SQL12 & 
- & 
Database &
7215 &
286 &
3.96 \\
\hline
 & 
 
 & \textbf{MEDIAN}
 &
\textbf{5683} &
\textbf{271} &
\textbf{4.77} \\\hline
\end{tabular}

\end{table}
\subsection{Data}

Table~\ref{table dataset details} shows the data used in this study.
This data comes  from the same source as Huang et al.; i,e the
publicly available dataset from Maldonado and Shihab~\cite{maldonado2015detecting}.
This dataset contains ten open source JAVA projects on different application domains, varying in size and number of developers and most importantly, in number of comments in source code. 
The provided dataset contains project names, classification type (if any) with actual comments. 
Note that, our problem do not concern with the type of SATD, rather we care about a binary problem of being a SATD or not. 
So, we have changed the final label into a binary problem by defining $\mathit{WITHOUT\_CLASSIFICATION}$ as $no$ and the rest (for example $DESIGN$) as $yes$.

When creating this dataset, Maldonado et al.~\cite{maldonado2015detecting} used jDeodrant~\cite{fokaefs2011jdeodorant}, an Eclipse plugin for extracting comments from the source code of java files. 
After that, they applied four filtering heuristics to the comments. 
A short description of them are given below (and for more details, see~\cite{maldonado2015detecting}):

\begin{itemize}
    \item Removed licensed comments, auto generated comments etc because according to the authors, they do not contain SATD by developers.
    \item Removed commented source codes as commented source codes do not contain any SATD.
    \item Removed Javadoc comments that do not contain  words like ``todo'', ``fixme'', ``xxx'' etc because according to the authors, the rest of the comments rarely contain SATD.
    \item Multiple single line comments are grouped into a single comment because they all convey a single message and it is easy to consider them as a group.
\end{itemize}

After applying these filters, the number of comments in each project reduced significantly (for example, the number of comments in Apache Ant reduced from $21,587$ to $4,140$, almost 19\% of the original size).

Two of the authors~\cite{maldonado2015detecting} then manually labelled each comments according to the six different types of TD mentioned by Alves et al.~\cite{alves2014towards}.
Note that if those labels were perfect, then {\IT}0 would
not be necessary.

\subsection{Standard Rig}

In the following, when we say {\em standard rig} we mean a 10-by-10 cross validation study that tries to build a predictor
for technical debt, as follows:
\bi
\item For i = 1 to 10 projects
\bi
\item test = project[i]
\item train = projects - test
\item 10 times repeat
\bi
\item Generate a new random number as seed. 
\item Apply the classifier $C$.
\bi
\item
For ensembles, we generate n-1 decision trees using the seed (learning from 90\% of the training data, selected at random).
\item For SVM, we shuffle the data using the seed. 
\ei
\item Apply {\IT}0, with $m=100$, stopping at some target recall
(usualy, 90\% recall for SATD). 
\ei
\ei
\ei
Note also, 
 when generating the estimator, we shuffle the data using the seed for building the logistic regression model.

\begin{table}[!b]
\centering
\caption{RQ1 results:
Agreement between 
{\IT}0's labels and those
generated by other methods.
Sorted by 100-recall. In this table, the {\em smaller}
the values in the right-hand-side column, the {\em larger}
the agreement between {\IT}0's labels  and other labels.}
\label{table ensemble dt classifier}
\begin{tabular}{|c|c|c|}
\hline
\textbf{Projects} & 
\textbf{Recall} & \textbf{100-Recall}\\
\hline
jruby    & 
88 & 12 \\ \hline
argouml  & 
87 & 13 \\ \hline 
columba  & 
87 & 13 \\  \hline
jmeter   & 
73 & 27 \\ \hline
hibernate & 
72 & 28 \\ \hline 
jfreechart & 
46 & 54 \\ \hline
emf      & 
33 & 67 \\ \hline 
sql12    & 
54 & 69 \\  \hline  
ant      & 
24 & 76 \\ \hline
jedit    & 
21 & 79 \\ \hline
\textbf{MEDIAN} & 
63 & 37 \\
\hline
\end{tabular}

\end{table}

\section{Results}
\label{section results}

The experimental materials described above where used to answer the 
  research questions from the introduction.

\subsection{RQ1: Is surveying necessary?}
 
Table~\ref{table ensemble dt classifier} reports the levels of (dis)agreement 
seen between the labels seen after rechecking and revising
labels (using {\IT}0) and the labels in the original data
This data was generated using our standard rig:
\bi
\item
In that table, we measure {\em disagreement}
as 100-recall.
\item
A   disagreement of 0\% indicates that the labels found via {\IT}0 are the same as in the original
data sets. 
\item
Note that the disagreements are quite large and range from 36\% to 79\% (median to max). That is: 
\ei
\begin{result}{1}
Surveying is required to resolve
 disagreement about labels.
\end{result}
When discussing this results with colleagues, they comment ``does
not that mean that {\IT}0 is just getting it wrong all the time?''.
We would argue against that interpretation. As shown below,
surveying improves
classification predictions  so whatever {\IT}0 is doing,
it is also improving the correspondence between the labels and the target concept.

Now suppose the reader is unconvinced by the last paragraph and wants to check whose labels are correct:
\bi
\item
The pre-existing labels?
\item
Or the labels generated by {\IT}0?
\ei
At that point the reader would
encounter the ``ground truth'' problems. That is, to assess
which labels are correct, the reader would need some ``correct'' labels. After some reflection (and a review of Table~\ref{tbl:costs}) the reader might realize that finding  those correct
set of labels can be very costly-- so much so that they would like some intelligent
assistant to help them label the data is a cost-effective manner. That is, they need
some tool like {\IT}0.

This is not a fanciful scenario. We envisage that once tools like {\IT}0 become
widespread, then informally ``labelling'' collectives will emerge between  collaborating 
research groups. Data sets would be passed between research groups, each one checking the labels
of the other. If the level  of disagreement on the next round of labelling
falls below a community-decided level of acceptability, then that data could then
move on to be used in research papers.



\subsection{RQ2: Is {\IT}0 useful?}
\begin{figure}[!t]
\begin{center}
\includegraphics[width=\linewidth]{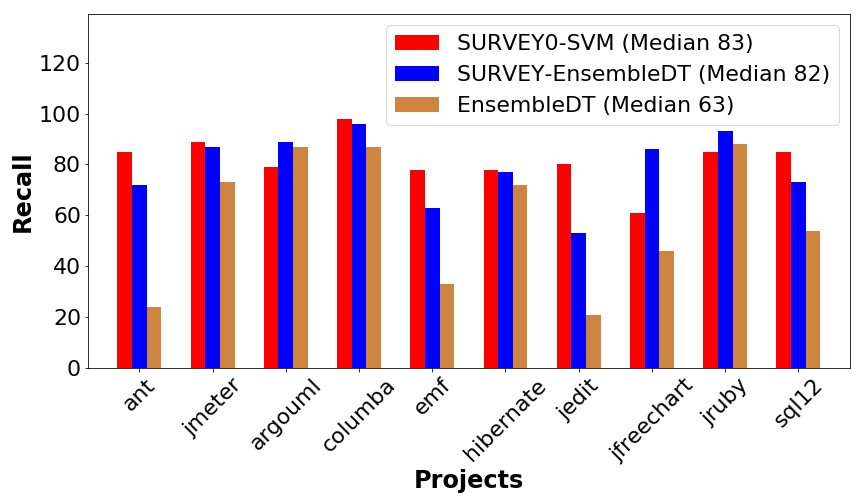}
\caption{Recall of {\IT}0, {\IT}-EnsembleDT and recall of Ensemble DT without {\IT}ing (from RQ1)}
\label{fig rq2 survey recall}
\end{center}
\end{figure}

\begin{figure}[!t]
\begin{center}
\includegraphics[width=\linewidth]{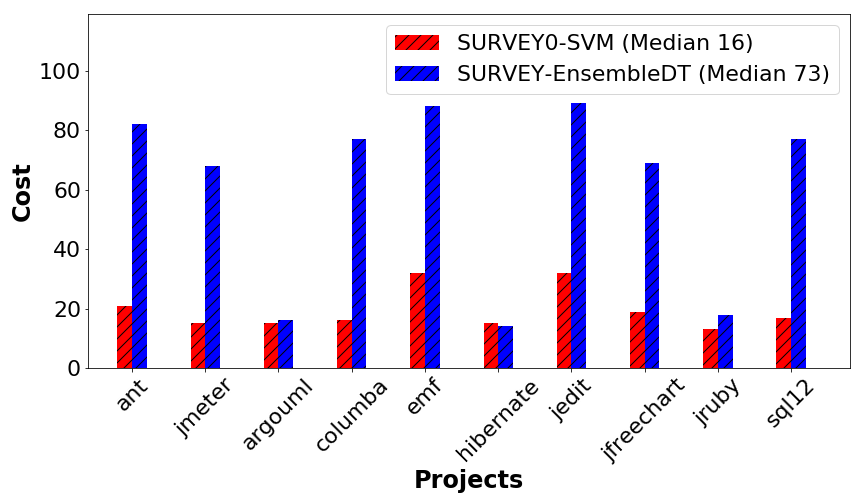}
\caption{Cost of {\IT}0 and {\IT}-EnsembleDT}
\label{fig rq2 survey cost}
\end{center}
\end{figure}

Figure~\ref{fig rq2 survey recall}
and 
Figure~\ref{fig rq2 survey cost} shows the recalls and costs achieved from the standard rig 
(when the target goal is 90\%). In those figures
\bi
\item The {\em EnembleDT} results are the closest we can come to reproducing
the methods of Huang et al. from EMSE'18. In these results, some classifier is learned
from nine projects, then applied to the tenth. These results make no use of {\IT}0;
i.e. here, there is no label review or revision. 
\item
The other plots come from {\IT0} using either EnembleDT or Linear SVM as the learner.
\ei
We observe that :
\bi
\item
The  two sets of treatments have median recalls of 82.5 and 62\% respectively.
\item
That is,
the treatments using {\IT}0 perform much better than those that do not.
\ei
That is:
\begin{result}{2}
{\IT}0 improved quality predictions. 
\end{result}
As to which classifiers we would recommend, Figure~\ref{fig rq2 survey recall}
reports that in terms of recall, both Linear SVM and EnsembleDT
perform just as well. However, Figure~\ref{fig rq2 survey cost} 
reports that Linear SVM has much lower associated cost; i.e. it can find
the technical debt comments much faster than EnsembleDT.

Based on these results, we  recommend {\IT}0
using a Linear SVM classifier

\subsection{RQ3: Is  {\IT}0 comparable to the state-of-the-art for human-in-the-loop AI?}

Certain   information retrieval methods offer   stopping criteria
for when to halt exploring new data. 
Here, we assess two such state-of-the-art approaches, developed
for assisting  Systematic Literature Reviews.

Those stopping  methods require
certain tuning parameters which we set by ``cheating''; i.e.  manually
tuning using  our test data. That is, we gave
the information retrieval methods an undue advantage over {\IT0}

  Ros et al.~\cite{ros2017machine}  suggests that, if no target class is found in $x$ consecutive examples seen (if each iteration offers $m$ examples, then a total of $x/m$ iterations), then we should stop.
Ros proposed $x=50$ but after ``cheating'',    we found
that    $x=10$ worked better (i.e. obtained higher recalls with minimum cost).

  Cormack et al.~\cite{cormack2016engineering}
finds the  knee in the current retrieval curve at each iteration and if the ratio between slops from $slope_{<knee}$ and $slope_{>knee}$ is greater than a predefined threshold $\rho$, then it stops. 
Note that, knee can be found using the \textit{Kneedle} algorithm~\cite{satopaa2011finding}. 
Cormack proposed  $\rho=6$
but after ``cheating'' we found that  $\rho=12$ was   a better value.

\begin{figure}[!t]
\begin{center}
\includegraphics[width=\linewidth]{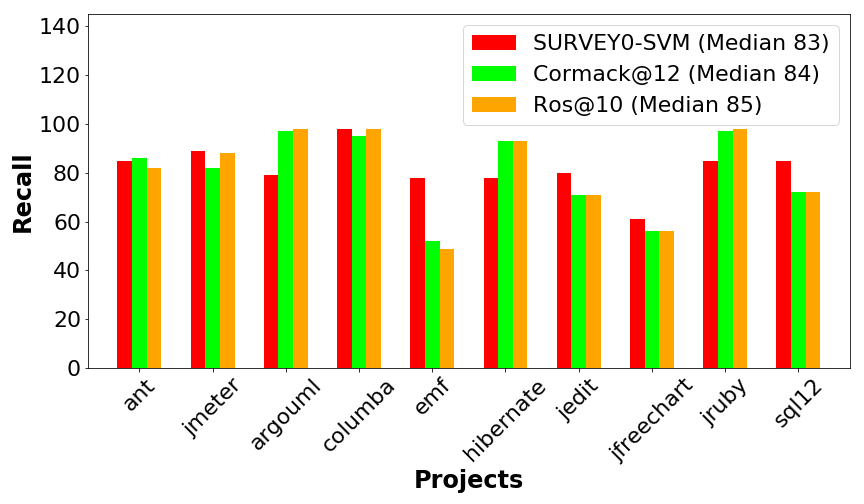}
\caption{Comparison between optimized state-of-the-art human-in-the-loop frameworks with {\IT}0}
\label{fig rq2 stopping recall}
\end{center}
\end{figure}

We compared these two baselines with our our standard rig.
As we can see from Figure~\ref{fig rq2 stopping recall}, even after
letting the other methods ``cheat'' (i.e. manually tuning these methods using the test data)  {\IT}0 wins on 6 projects out of 9 and has an overall recall almost as good as the   ``cheating'' results. 
Thus, we   say:

\begin{result}{3}
{\IT}0 made its predictions
at a near optimum rate.
\end{result}

{\bf RQ4: How soon can {\IT}0 learn quality predictors?}
As we know from RQ2 and RQ3, that {\IT}0 has a high recall, meaning, when looking for SATDs, it finds most of them. But in order to do that, {\IT}0 need a human expert to read through the comments suggested by the classifier and sorter. Thus, a core part of {\IT}0 is to ensure that, the cost of reading is minimized and learning when to stop. 

Our experiment with {\IT}0 on Target@90 recall show that, after reading only 16\% (median) of the comments, {\IT}0 stops while finding 83\% (median) of the SATDs. This 16\% cost has an IQR\footnote{IQR= inter-quartile-range = (75-25)th percentile.} of 5\% across project, implying, the cost is nearly   the same for all ten projects. 
Hence, we say:
\begin{result}{4}
{\IT}0 can be recommended as a way to reduce the labelling effort.
\end{result}

{\bf RQ5: Can {\IT}0 find more     issues?}

In the above experiments, we set that target goal to be 90\% recall.
Here, we report what happens when we seek to find more technical debt;
i.e when we set the target recall to 95\%.

Table~\ref{table rq5 90 95} shows the results.
As before, if we set the target to X\%, we achieve
a performance level of slightly less than X
(so the median recalls achieved when the target was 90\% or 95\%
was 83\% or 89\%, respectively).

We also see that increasing the target recall by just 5\% (from 90 to 95) nearly doubles the cost of finding the technical debt
(from reading 16\% of the comments to 29\%). 
We make no comment here on whether or not it is worth
increasing the cost in this way. All we say is that, if
required, our methods can be used to tune how much work
is done to reach some deseired level of recall. That is:
\begin{table}[!t]
\centering
\caption{RQ5: Cost Effective {\IT}0 for Target@90 and Target@95.}
\label{table rq5 90 95}
\begin{tabular}{|c||c|c|c|c|}
\hline
\multicolumn{1}{|c||}{\textbf{Projects}} & 
\multicolumn{2}{c|}{\textbf{Target@90}} &
\multicolumn{2}{c|}{\textbf{Target@95}} \\ 
\hline
\textbf{Name} & 
\textbf{Recall} & \textbf{Cost} &
\textbf{Recall} & \textbf{Cost} \\
\hline
ant  & 
85 & 21 &
87 & 35 \\
\hline
jmeter & 
89 & 15 &
93 & 38 \\
\hline
argouml & 
79 & 15 &
90 & 21 \\
\hline
columba & 
98 & 16 &
98 & 33 \\
\hline
emf & 
78 & 32 &
83 & 41 \\
\hline
hibernate & 
78 & 15 &
85 & 21 \\
\hline
jedit & 
80 & 32 &
86 & 44 \\
\hline
jfreechart & 
61 & 19 &
66 & 20 \\
\hline
jruby & 
85 & 13 &
94 & 19 \\
\hline
sql12 & 
85 & 17 &
91 & 24 \\
\hline
\hline
\textbf{MEDIAN} & 
\textbf{83} & \textbf{16} &
\textbf{89} & \textbf{29} \\
\hline
\textbf{IQR} & 
7 & 5 &
7 & 16 \\
\hline
\end{tabular}
\end{table}

\begin{result}{5}
 {\IT}0 can be used to   advise
on   how much more work is required to achieve 
some additional desired level of quality assurance.
\end{result}

\subsection{ RQ6: How much does {\IT}0 delay human readers?}
{\IT}0 need a human expert in the loop to classify the most possible datapoint. To that end, {\IT}0 offer $m$ examples at each iteration, before estimating the remaining target class (here SATD). 

This estimation process has its own overhead. According to Maldonado et al., each example need approximately 10.3 seconds to classify. So, if $m$ examples are offered at each iteration, then the human expert will need approximately $m \times 10.3$ seconds to finish reading. If the estimation process takes longer than this, then human expert become unproductive while waiting for the next iteration. After experimenting, we see that on average, the estimation process takes 30 seconds for $m=100$. On the other hand, for $m=100$, human reader will need $100*10.3$ seconds or approximately 20 minutes. If we calculate the overhead of each iteration, it is only $0.5/(20+0.5) \approx 2\%$. To put that another way:
\begin{result}{6}
{\IT}0 imposed
negligible overhead  (i.e. less than 5\%) on the activity of human experts.
\end{result}



\section{Threats to Validity}
\label{section threats to validity}
 


\textbf{Model Bias:} One internal threat to validity is our bias towards the classifier selection and stopping rule selection. 
We experimented a wide variety of state-of-the-art classifiers used in text-mining as rankers while building {\IT}0 and found SVM to be the best. 
Yet, there are other advanced and complex classifiers (such as LSTM) that we did not used in our selection, because of the simplicity of our dataset as well as no prior work has used them to classify SATDs. 
We also avoided a few stopping rules as baselines (such as Wallace~\cite{wallace2013active}) intentionally as previous research showed~\cite{yu2017fast} that our baselines are significantly better than theirs.
Nevertheless, we
 are aware that our model selection is not comprehensive and could be explored further in future research.

\textbf{Evaluation Bias:} We have reported Recall and Cost as our overall measure.
We have repeated each experiment ten times and reported only the median values for minimizing any bias towards randomness. 
We understand the quality measures are not comprehensive and their might be other quality measure used in software engineering that reflects more comprehensive summary of our findings. 
A more comprehensive analysis using other measures is left for future work. 

\textbf{Sample Bias:} The dataset was provided by authors Maldonado and Shihab~\cite{maldonado2015detecting}. Other data might lead
to other conclusions. Clearly, this work needs to be repeated
on other data.




\section{Discussion}
\label{section discussion}

In our work, we have studied the comments of ten open source projects developed in JAVA. 
Our work shows that, with minimum cost, we can identify
self-admitted technical debt
using a combination of AI with human. 
There are several ways to extend the current work.

\begin{itemize}
    \item \textit{Feature Selection and Vectorization:} A few recent work shows that feature selection can improve the overall classification of SATD~\cite{huang2018identifying}. 
    A more recent work also implies that word embedding model such as word2vec is promising while identifying SATD~\cite{flisar2018enhanced}. 
    We believe, our framework can also improve significantly after proper feature selection and vectorization. 
    We initially did some feature selection, but a more rigorous experiment must be done in this regard in the future.
    \item \textit{New Dataset:} Our work is confined in Java projects and Open Source Projects. 
    We want to develop new dataset to generalize our findings and possibly discover new facts along the way.
    \item \textit{Matrices:} There are other goal metrics
    to explore.  
    For example, measuring cost in terms of time or man-hour might be a better quality measure. 
    \item \textit{Results:} According to our experiment, we can find $83\%$ SATD while only reading $16\%$ of the data (both in median).
    We hypothesize that, this results can be improved using hyperparameter optimization. 
    The only drawback is the run-time of such tuning. In our future work, we will try to find improved results using hyperparameter tuning. 
\end{itemize}

\section{Conclusion}
Technical debt is a metaphor to describe the quick and dirty workaround to receive immediate gain. 
This is an intentional practice and often developers leave intentional comments indicating that their work is sub-otimal. 
Although this phenomena is unavoidable in reality, research show that the long term impact of these practice is dire. 
Thus identifying technical debt is a major concern for the industry.
This work has explored methods for building a technical debt predictor, at minimal cost.

The methods used here to reduce the cost of building technical debt
predictors are quite general to any human-in-the-loop process where some subject matter expert is required to read and label a large corpus. Such work can be time-consuming, tedious, and error-prone.
Our work is a response to that. 
We offer a complete framework where human will be guided by an AI to label artifacts with minimal effort. 
At least for the data studied here,
we can find on average (median) $83\%$ of the artifacts
of interest   by reading only $16\%$ of the artifacts. 
Examining the possible implication on a larger dataset with better estimator and well tuned parameters will open interesting possibilities in future. 

\bibliographystyle{IEEEtran}
\balance
\bibliography{bibliography}

\end{document}